# Impact of Se concentration and distribution on topological transition in FeTe1-xSex crystals

Jinying Wang[1*], Gerhard Klimeck[1-2]

1. Network for Computational Nanotechnology, Purdue University, West Lafayette, Indiana 47907, USA

2. School of Electrical and Computer Engineering, Purdue University, West Lafayette, Indiana 47907, USA

* corresponding authors: wang4205@purdue.edu

# Abstract

A topological transition in high-temperature superconductors $FeTe_{1-x}Se_x$, occurring at a critical range of Se concentration $x$, underlies their intrinsic topological superconductivity and emergence of Majorana states within vortices. Nonetheless, the influence of Se concentration and distribution on the electronic states in $FeTe_{1-x}Se_x$ remains unclear, particularly concerning their relationship with the presence or absence of Majorana states. In this study, we combine density functional theory (DFT) calculations, $p_z$-$d_{xz/yz}$-based and Wannier-based Hamiltonian analysis to systematically explore the electronic structures of diverse $FeTe_{1-x}Se_x$ compositions. Our investigation reveals a nonlinear variation of the spin-orbit coupling (SOC) gap between $p_z$ and $d_{xz/yz}$ bands in response to $x$, with the maximum gap occurring at $x = 0.5$. The $p_z$-$p_z$ and $d_{x2-y2}$-$p_z$ interactions are found to be critical for pd band inversion. Furthermore, we ascertain that the distribution of Se significantly modulates the SOC gap, thereby influencing the presence or absence of Majorana states within local vortices.

# Introduction

The iron chalcogenide family has emerged as an intriguing platform to study unconventional superconductivity and novel quantum state transitions[1–3]. In particular, the recent discovery of topological superconductivity and Majorana bound states (MBS) in $FeTe_{0.55}Se_{0.45}$[4,5] has significantly promoted topological quantum computing research and has obtained immense attention. The unique electronic properties of this system have been suggested to depend strongly on the Te/Se concentration both theoretically[6,7] and experimentally[8]. In contrast to the topologically trivial FeSe, $FeTe_{1-x}Se_x$ ($x = 0.5$ in first-principles calculations[7] and x = 0.45 in experiments[5]) exhibits nontrivial topological surface states (TSS) due to the parity inversion and spin-orbit coupling (SOC) gap opening of p and d bands along the $\Gamma$ to Z point in the Brillouin zone, induced by Te substitution. The sensitivity of topology in iron chalcogenide to chemical composition is also supported by recent angle resolved photoemission spectroscopy (ARPES) measurements[8], as TSS appears only at sufficiently high Te concentration, and $FeTe_{0.55}Se_{0.45}$ locates near the boundary among non-superconducting, normal superconducting, and topological superconducting phases. The coexistence of TSS and superconductivity gives rise to MBS in vortices, which can be detected using scanning tunneling microscopy (STM) and spectroscopy[4,5]. However, not all vortices host MBS[9]. The reasons behind the presence/absence of MBS have been intensely debated, including vortex overlap[9], vortex disorder[10], Zeeman coupling[11], and local composition fluctuations[12], etc. Based on an effective Hamiltonian analysis, Sau et al.[12] have revealed that local fluctuations in Se concentration lead to chemical potential disorders as well as topological domain disorders, thus influencing the presence of MBS. Additionally, local composition fluctuations induce local strains, resulting in nematic transitions and nanoscale suppression of superconductivity in $FeTe_{1-x}Se_x$[13]. Despite previous studies highlighting the significance of Se concentration for TSS and MBS, important questions such as the critical Se concentration for TSS and its impact on the electronic states remain unclear.

Exploring the chemical composition effect on the electronic states of $FeTe_{1-x}Se_x$ poses considerable challenges from both experimental and theoretical perspectives. High-quality growth of $FeTe_{1-x}Se_x$ crystals with precise composition control remains a difficult task. The atomic structures are highly sensitive to the Se/Te concentration, and there are limited reports available[8,14,15], making it even more challenging to unveil their atomic structures. The presence of local disorders further complicates the situation[8]. Although density functional theory (DFT) can offer qualitatively correct explanations of electronic states for certain compositions (e.g., $x = 0$, 0.5, 1)[7], its failure to accurately describe strong correlation and coupling effects leads to large band renormalization factors (R) when compared to angle-resolved photoemission spectroscopy (ARPES) data[16]. Additionally, these band renormalization factors are dependent on the composition and orbital characteristics[6]. Alternative methods such as DFT+U and dynamical mean-field theory (DMFT) can improve the renormalization, but they still fail to reproduce the experimental band values[6,17]. Effective Hamiltonian studies[16,18] mainly rely on experimental or first-principles results, and lack a direct correlation between composition and Hamiltonian terms. Currently, our understanding of the composition effect on orbital interactions and electronic states in $FeTe_{1-x}Se_x$ with varying $x$ is deficient even at a comprehensive qualitative level, let alone achieving an accurate quantitative description.

Here, we focus on studying the relation among chemical composition, orbital interaction, and electronic structures. By combining DFT calculations, Wannier-based TB, and a $p_z$-$d_{xz/yz}$-based effective Hamiltonian analysis together, we systematically study the chemical composition effect on electronic states of FTS, especially the bands variation along $\Gamma$-Z around Fermi level. The dependence of orbital interaction with x are clarified.

# Models

Variation of Se concentration $x$ in FeTe$_{1-x}$Se$_x$ crystals changes not only Se/Te proportion but also atomic structures. The structures of FTS are limited reported and slightly vary with preparation and measurement methods[19–22]. The lattice constants $a$ and $c$ in FTS gradually decrease by about ~0.05 Å (~2 %) and ~0.76 Å (~12 %), respectively, when $x$ increases from 0 to 1. The relation between $a/c$ and $x$ can be approximately in linear ways[22]. The bond lengths between Fe and Se/Te atoms or heights between Fe and Se/Te planes ($d_z$) are fewer reported[7,21], and the $d_z$-$x$ relation displays more complex than the $a/c$-$x$ linear relations. Besides, uneven local distribution of Se/Te atoms generally exists in FeTe$_{1-x}$Se$_x$[9]. Here, we mainly study the average composition effect on band structures of FeTe$_{1-x}$Se$_x$ ($x$ = 0.0, 0.125, 0.25, 0.375, 0.5, 0.625, 0.75, 0.875, 1.0, see in Fig. 1a) systems using 2*2*1 supercells of primitive FeTe$_{0.5}$Se$_{0.5}$ crystal structure[7]. Because the electronic structures of FeTe$_{1-x}$Se$_x$ are quite sensitive to both Se/Te proportion and crystal structures, we adopt experimental-based linear variation of lattice constants with Se concentrations: $a$ = 7.6406 - 0.108$x$ (Å), $c$ = 6.3362 - 0.762$x$ (Å), and optimize the atomic geometry using DFT calculations implemented in VASP.

# Methods

Although DMFT and other beyond DFT methods give better description of correlation and coupling effects, all these calculations show $x$-dependent and orbital-dependent band renormalzation compared to ARPES data[6]. Instead of seeking for accurate band calculations, we focus on studying the critical orbital interactions caused by x changes by combining DFT calculations, Wannier TB, and a $p_z$-$d_{xz/yz}$-based Hamiltonian analysis together.

1. DFT

DFT calculations with GGA-PBE[23] functionals implemented in VASP[24,25] are used to study the electronic structures of FeTe$_{1-x}$Se$_x$ crystals. Van-der Walls interactions are included in the optimization of FeTe$_{1-x}$Se$_x$ systems by using vdW-DF2 functional[26,27] after many tests. The cutoff energy is set to be 400 eV, and (7 × 7 × 15) Monkhorst-Pack grids are used. We set the Fermi level to be zero in all cases.

2. Wannier TB

Tight-binding (TB) Hamiltonians based on maximally localized Wannier functions of FeTe$_{1-x}$Se$_x$ systems are constructed using Wannier90[28]. A basis set of 5d orbitals ($d_{xy}$, $d_{xz}$, $d_{yz}$, $d_{z2}$, $d_{x2-y2}$) of each Fe atom and 3p orbitals ($p_x$, $p_y$, $p_z$) of each Se/Te atom are used here. The obtained Wannier functions show the same features with corresponding atomistic orbitals.

3. $p_z$-$d_{xz/yz}$-based Hamiltonian

A $p_z$-$d_{xz/yz}$-based Hamiltonian[16] is used to analyze the band dispersion along Γ-Z direction of FeTe$_{1-x}$Se$_x$ systems with and without SOC effect,

$$H(k_z) = \begin{pmatrix} \varepsilon_p + 2t_{pp}\cos k_z & -2\lambda_3\sigma_x\sin k_z & 2\lambda_3\sigma_y\sin k_z \\ . & \varepsilon_d + 2t_{dd}\cos k_z & i\sigma_z(\lambda_1 + 2\lambda_2\sin k_z) \\ . & . & \varepsilon_d + 2t_{dd}\cos k_z \end{pmatrix} \qquad (1)$$

# Main results and discussion

The optimized atomic structures of FeTe$_{1-x}$Se$_x$ ($x$ = 0.0, 0.125, 0.25, 0.375, 0.5, 0.625, 0.75, 0.875, and 1.0, respectively) are shown in Fig. 1a, where the Fe-Se and Fe-Te bond lengths are 2.34 Å and 2.51 Å, respectively. The Fe, Se, and Te coordinates remain stable in proximity to their original lattice positions.

Neither obvious atomic distortion nor rearrangement is found when $x$ changes. Fig. 1b illustrates three distinctive bands near the Fermi level, which are primarily composed of $p_z$ orbitals of chalcogen and $d_{xz/yz}$ and $d_{x2-y2}$ orbitals of Fe according to projected density of states analysis. The $p_z$ and $d_{xz/yz}$ bands along $\Gamma$ to

Z has shown to be critical for the topology in FeSe$_{0.5}$Te$_{0.5}$[7,18]. As the Se concentration ($x$) gradually increases from 0 to 1, the $p_z$ band at $\Gamma$ exhibits a noticeable upward shift, ultimately leading to the vanishing of band inversion between $p_z$ and $d_{xz/yz}$ bands when $x$ closes to 1. The SOC effect not only causes a split in the degenerate $d_{xz/yz}$ bands but also facilitates a hybridization between $p_z$ and $d_{xz/yz}$ bands, thereby opening an

SOC gap between them along the $\Gamma-Z$ path. This SOC gap signifies a topological transition within FeTe$_{1-}$

$_x$Se$_x$ systems, which further leads to the manifestation of topological superconductivity and MBS within

vortices at sufficiently low temperatures[4,5].

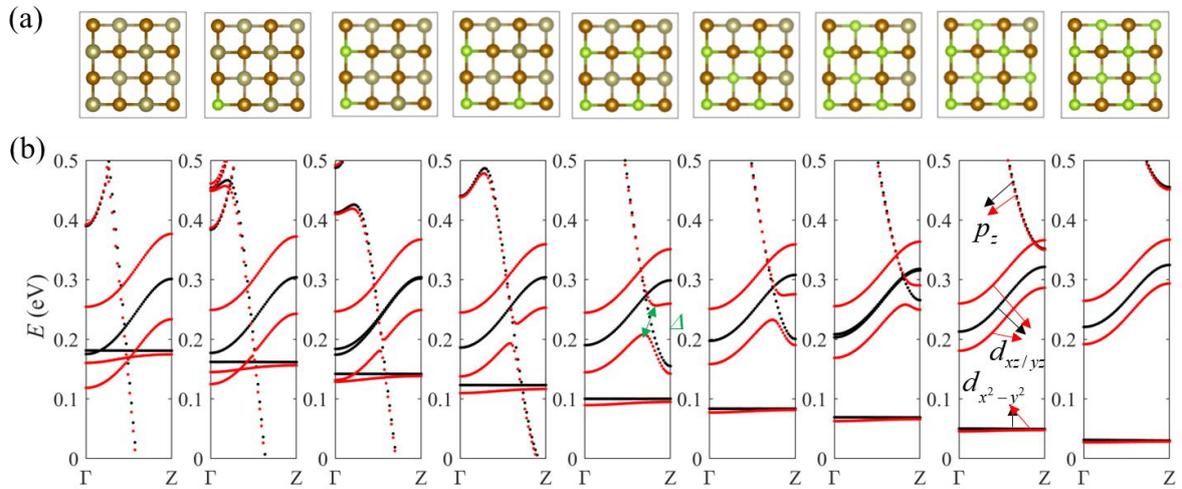

Figure 1. (a) Atomic structures and (b) band structures of FeTe$_{1-x}$Se$_x$ ($x$ = 0.0, 0.125, 0.25, 0.375, 0.5, 0.625,

0.75, 0.875, 1.0) calculated by DFT implemented in VASP with (red dot) and without (black dot) SOC effect.

The approximated linear relation between lattice constants and $x$ for FeSe$_x$Te$_{1-x}$ based on experimental data[]

is used, that is, $a$ = 7.6406 - 0.108$x$ (Å) and $c$ = 6.3362 - 0.762$x$ (Å). Atomic positions are optimized using revised Perdew-Burke-Ernzerhof functional[26] with nonlocal vdW-DF2 functional[27].

Interestingly, as $x$ progresses from 0 to 1, the magnitude of the SOC gap experiences a nonlinear evolution: it begins at 0.0 meV, reaches its maximum value of 48.2 meV at $x = 0.5$, and subsequently recedes back to 0.0 (see Fig. 2a). This distinctive nonlinearity in the variation of SOC gap with $x$ is found to be unaffected by optimization methods or lattice structures, hinging solely on the change of chemical composition (see Fig. S1-S2 in supplementary materials). To be noted, larger SOC gap should correspond to more robust MBS.

To understand how Se concentration affects the orbital interactions and energy bands of FeTe$_{1-x}$Se$_x$, we adopt a $p_z$-$d_{xz/yz}$-based Hamiltonian[16] as well as Wannier-based TB approach with basis sets of 5d (Fe) and 3p (Se/Te) orbitals to fit and analyze our DFT results (see Fig. S3 in supplementary materials). The effective pd Hamiltonian reproduces the $\Gamma$−Z bands in excellent agreement with DFT results. The onsite and hopping energies, denoted as $\varepsilon_p$, $\varepsilon_d$, $t_{pp}$ and $t_{dd}$, mainly control the p and d bands shape and energy ranges. When $x$ increases from 0 to 1, we observe a gradual increase in both $\varepsilon_p$ and $\varepsilon_d$, accompanied by a decrement in $t_{pp}$ and small variation of $\varepsilon_d$ and $t_{dd}$ (see Fig. 2b). The inversion of p and d bands along $\Gamma$−Z path transpires when the signs of $(\varepsilon_p + 2t_{pp}) - (\varepsilon_d + 2t_{dd})$ and $(\varepsilon_p - 2t_{pp}) - (\varepsilon_d - 2t_{dd})$ differs. The splitting and curvature of $d_{xz/yz}$ bands are predominantly influenced by the $d_{xz}$-$d_{yz}$ interaction term $i\sigma_z(\lambda_1 + 2\lambda_2 \sin k_z)$ and its conjugate. The $p_z$-$d_{xz/yz}$ interaction which is a function of $\lambda_3$ mainly governs the SOC gap between $p_z$ and $d_{xz/yz}$ bands, mirroring a similar trend of variation with the SOC gap itself (see Fig. 2a).

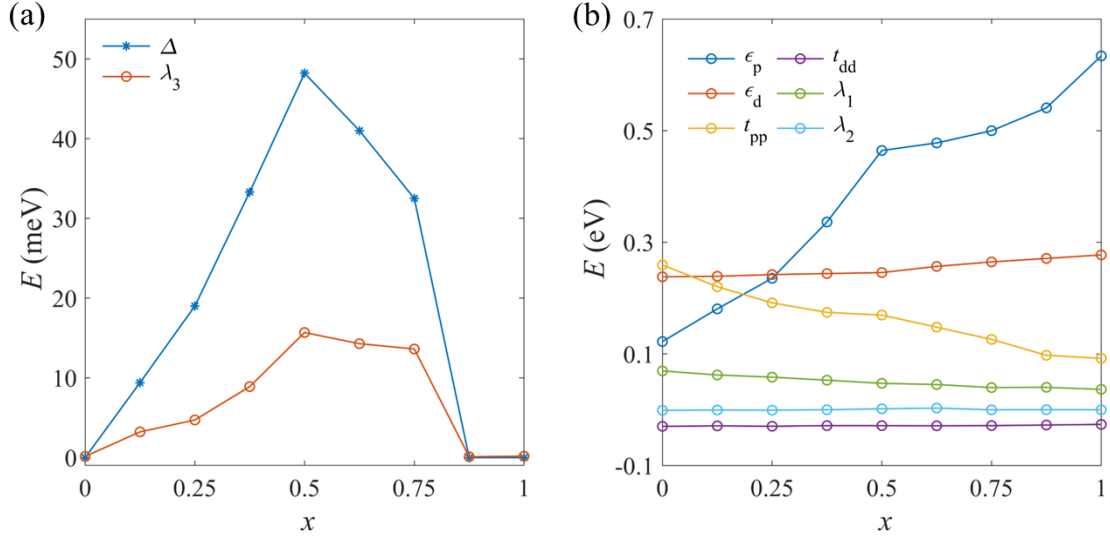

Figure 2. (a) DFT calculated SOC gap $\Delta$ and fitted $\lambda_3$ in effective pd Hamiltonian as well as (b) other pd Hamiltonian parameters of FeTe$_{1-x}$Se$_x$ ($x$ = 0.0, 0.125, 0.25, 0.375, 0.5, 0.625, 0.75, 0.875, 1.0).

We extend our analysis by employing a Wannier-based TB approach for FeTe$_{1-x}$Se$_x$ ($x$ = 0, 0.5 and 1) systems to elucidate the intricate correlation between Se concentration and atomic orbital interactions. The fitted Wannier Hamiltonian shows good description of DFT data (see Fig. S4 in supplementary materials), and the main interaction parameters governing the $p_z$, $d_{xz/yz}$, and $d_{x2-y2}$ bands are listed in Table 1. Some Wannier-based Hamiltonian parameters exhibit variations consistent with those revealed through effective pd Hamiltonian analysis. For example, the on-site energies of $d$ orbitals (($\varepsilon_{d_{xz/yz}}$, $\varepsilon_{d_{x^2-y^2}}$) vary little with $x$, while the $p$-$p$ hopping energies ($t^0_{pz\sim pz}$, $t^1_{pz\sim pz}$) decline as $x$. This observation mirrors the inherent stability of Fe lattices under varying $x$ and the greater localization of $p$ orbitals in Se compared to Te. The Wannier analysis reveals that the interaction between $p_z$ and $d_{x2-y2}$ is critical for $p_z$-$d_{xz/yz}$ band inversion, which is ignored in the effective pd Hamiltonian. A stepwise substitution of Wannier Hamiltonian parameters from $x$ = 1 with those from $x$ = 0 (see Fig. S5 in supplementary materials) vividly demonstrates that even a slight reduction in $\left|t_{dx^2-y^2 \sim p_z}\right|$ triggers a significant descending of $p_z$ band. Consequently, the initially separate $p_z$ and $d_{xz/yz}$ bands intersect and undergo a band inversion. Although other Hamiltonian parameters change with $x$, their impact on $p_z$-$d_{xz/yz}$ band inversion remains modest, except for the intralayer $p_z$-$p_z$ interaction $t^0_{pz \sim pz}$ which not only shifts the pz band to lower energy range but also broadens it. In summary, $\left|t_{dx^2-y^2 \sim p_z}\right|$ and $t^0_{pz \sim pz}$ emerge as the most important orbital interactions which mainly determines the occurrence of pd band inversion. In contrast to the evident impact of $p_z$-$d_{xz/yz}$ interactions on SOC gap in eq. (1), multiple interactions encompassing pz and d orbitals contribute to the SOC gap in Wannier Hamiltonian. Nonetheless, it remains uncertain how these interactions are influenced by $x$ and subsequently govern the SOC gap from the point of the Wannier-based atomistic-orbital perspective.

Table 1. Main Hamiltonian parameters in Wannier-functions-based Hamiltonians for FeSe$_x$Te$_{1-x}$ ($x$ = 0.0, 0.5, 1.0) systems without and with SOC effect. The on-site energies of $d_{xz/yz}$, $d_{x2-y2}$ orbitals of Fe and $p_z$ orbitals of chalcogen are noted as $\varepsilon_{d_{xz/yz}}$ , $\varepsilon_{d_{x^2-y^2}}$ , $\varepsilon_{p_z(Se,Te)}$, respectively. And $t_{p_z(Se,Te)-d_{x^2-y^2}}$ , $t_{p_z(Se,Te)-d_{xz/yz}}$, $t^0_{p_z-p_z}$ , $t^1_{p_z-p_z}$ are the hopping energies of nearest-neighbor $p_z$-$d_{x2-y2}$, $p_z$-$d_{xz/yz}$, intralayer $p_z$-$p_z$, and interlayer $p_z$-$p_z$ interaction terms, respectively. Fermi energy is set to be zero.

| | $x$ | 0.0 | 0.5 | 1.0 |
|---|---|---|---|---|
| $H_{W90}$ woSOC | $\varepsilon_{dxz/yz}$ (eV) | -0.694 | -0.733, -0.651 | -0.688 |
| | $\varepsilon_{dx2-y2}$ (eV) | -0.721 | -0.745 | -0.779 |
| | $\varepsilon_{pz(Se, Te)}$(eV) | -2.758 | -3.053, -2.679 | -2.976 |
| | $/t^0_{pz(Se, Te)-dx2-y2}|$ (eV) | 0.549 | 0.608, 0.640 | 0.672 |
| | $/t^0_{pz(Se,Te)-dxz/yz}|$ (eV) | 0.196 | 0.237, 0.092 | 0.164 |
| | $t^0_{pz-pz}$(eV) | 0.268 | 0.224 | 0.186 |
| | $t^1_{pz-pz}$(eV) | 0.322 | 0.318 | 0.302 |
| $H_{W90}$ wSOC | $\varepsilon_{dxz/yz}$(eV) | -0.720 | -0.717, -0.631 | -0.713 |
| | $\varepsilon_{dx2-y2}$(eV) | -0.786 | -0.773 | -0.848 |
| | $\varepsilon_{pz(Se, Te)}$(eV) | -2.778 | -3.083, -2.728 | -2.990 |
| | $/t^0_{pz(Se, Te)-dx2-y2}|$ (eV) | 0.548 | 0.606, 0.640 | 0.694 |

| | | | |
|---|---|---|---|
| $\|t^0_{\text{pz(Se,Te)-dxz/yz}}\|$ (eV) | 0.196 | 0.238, 0.092 | 0.143 |
| $t^0_{\text{pz-pz}}$(eV) | 0.267 | 0.223 | 0.186 |
| $t^1_{\text{pz-pz}}$(eV) | 0.322 | 0.318 | 0.303 |

The above discussion about the impact of Se concentration on the electronic structure of FeTe$_{1-x}$Se$_x$ is based on specific periodic structures for each $x$ value which maximize the dispersion of Se and Te atoms. However, real samples manifest uneven distributions of Se and Te atoms. For instance, STM mappings have indicated that the fluctuations in local Se concentration on surfaces surpass 10%[9]. To address the variability in the distributions of Se atoms, we first undertake calculations for FeTe$_{0.5}$Se$_{0.5}$ using differing atomic configurations achieved by modifying the positions of Se/Te within a 2*2 supercell.

As shown in Fig. 3, FeTe$_{0.5}$Se$_{0.5}$ with different Se distribution exhibit completely distinct electronic structures, with their SOC gaps ranging from 0.0 to 48.2 meV. Especially, the structure illustrated in Fig. 3c shows a SOC gap of 0 meV. Due to the overlapping of surface states and bulk states, the surface is topologically trivial, making the formation of MBS within vortices nearly implausible. Furthermore, we observe variations in the energy levels of SOC gap within FeTe$_{0.5}$Se$_{0.5}$ across different atomic distributions. Considering diverse local domains corresponding to these four atomic configurations (Fig. 1a and Fig. 3a-c) and placing the chemical potential at the marked black dash in Fig. 3d, it becomes apparent that only the local domain resembling the first atomic distribution (as seen in Fig. 1a with $x$ = 0.5) is primed for a surface phase transition towards topological superconductivity and the consequent formation of robust MBS, as the chemical potential locates in the SOC gap[18]. In contrast, the formation of MBS in the other domains is hindered by the substantial occupation of bulk states and the presence of topologically trivial surfaces, attributed to the chemical potential residing well above the SOC gap. Our calculation unveil that the uneven dispersion of Se/Te atoms leads to fluctuations in both SOC gap and chemical potential of local domains.

These fluctuations account for the presence and absence of MBS in vortices.

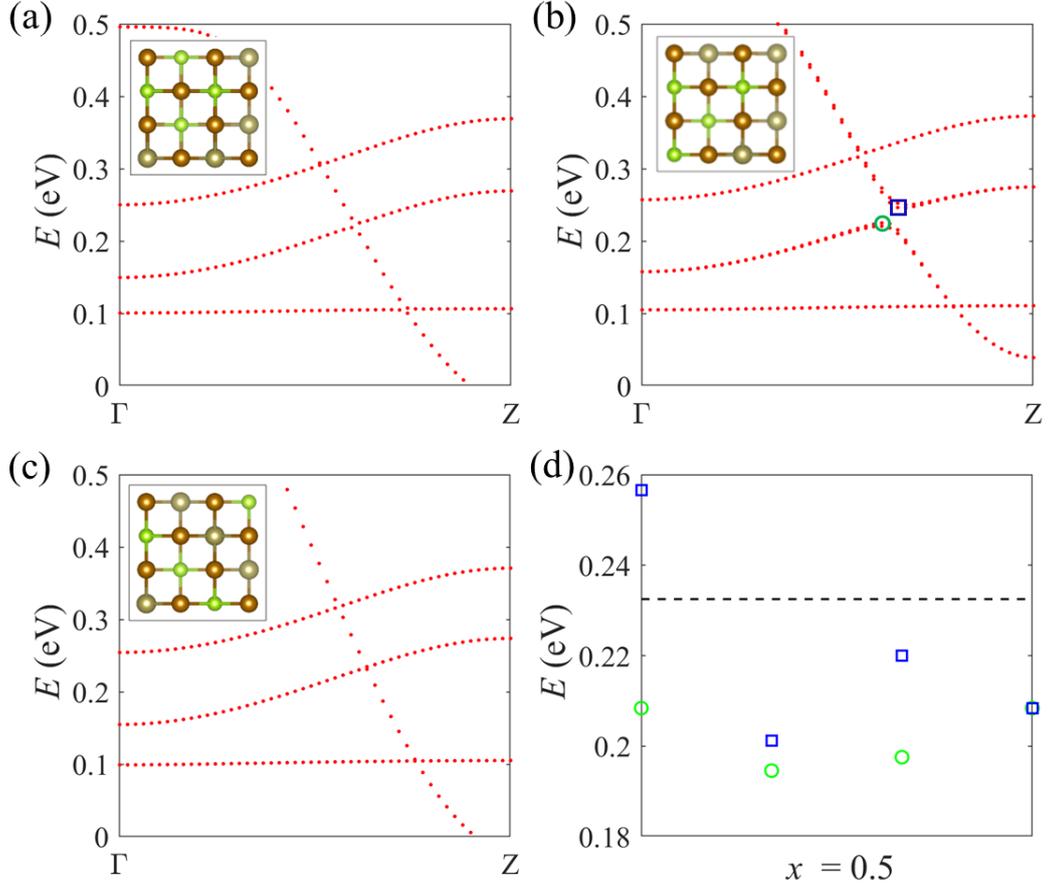

Figure 3. (a-c) Band structures of FeTe$_{0.5}$Se$_{0.5}$ with three different types of Se/Te atomic distribution. Insets are top views of the atomic structures. (d) The two energy points which determines SOC gap (blue square and green circle) for all four distinct atomic configurations of FeTe$_{0.5}$Se$_{0.5}$, and the chemical potential is marked as black dash line.

We further explore the electronic properties of the FeSe$_x$Te$_{1-x}$ across varying Se concentrations and distributions. The SOC gap exhibits a nonlinear and intricate behavior as $x$ changes (Fig. 4). As the local disorder of Se concentration and distribution increases, it can be expected that both SOC gap and its energy levels in local domains fluctuate. Considering the strong correlation between the SOC gap and chemical potential window for TSC phase in FeSe$_x$Te$_{1-x}$, the MBS tends to diminish as the SOC gap contracts. Previous studies have shown the influence of chemical composition on chemical potential and band inversion in FeSe$_x$Te$_{1-x}$ systems[7,12,18]. For instances, Sau et al. have emphasized the distinctive features between topological disorder arising from band inversion and chemical potential disorder as well as their impacts on the emergence of MBS when Se concentration changes[12]. However, our discovery of the non-linear and multi-scattering characteristics of Se concentration and distribution in relation to the SOC gap offers a fresh perspective to rationalize the appearance and disappearance of MBS in localized vortices. The establishment of the TSC phase and local MBS exclusively materializes when the chemical potential aligns with the local SOC gap; otherwise, the localized domains revert to a topologically trivial state.

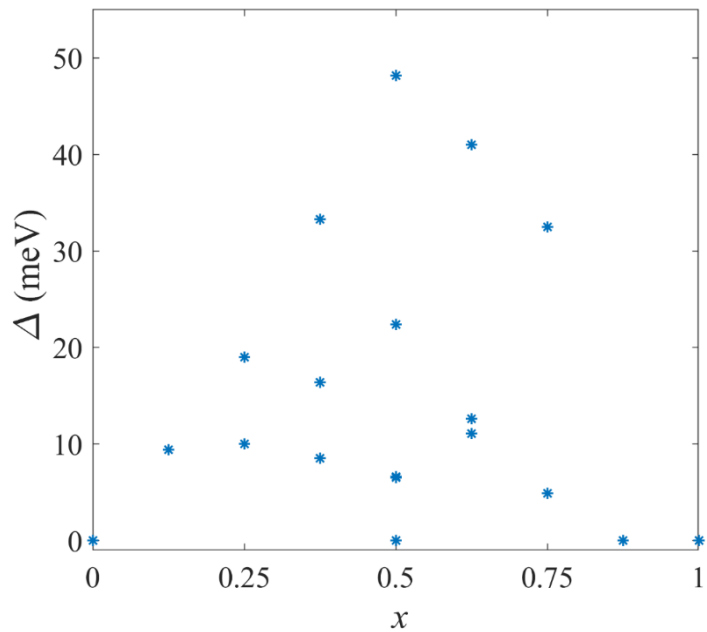

Figure 4. SOC gap of FeTe$_{1-x}$Se$_x$ ($x$ = 0.0, 0.125, 0.25, 0.375, 0.5, 0.625, 0.75, 0.875, 1.0) with different Se distributions.

# Conclusion

In conclusion, our study explores the intricate relationship between Se composition and electronic states in $FeTe_{1-x}Se_x$. By combining DFT calculations and Hamiltonian analysis, we uncover a nonlinear variation of the SOC gap between $p_z$ and $d_{xz/yz}$ bands in response to $x$, and find the critical role of $p_z$-$p_z$ and $d_{x2-y2}$-$p_z$ interactions in pd band inversion. Importantly, our findings underscore the substantial impact of Se distribution on the SOC gap, subsequently influencing the presence or absence of MBS within local vortices. This study reveals the intricate interplay between chemical composition, electronic structure, and Majorana in $FeTe_{1-x}Se_x$, providing a novel perspective to understand the appearance and disappearance of MBS.

# Acknowledgment

This work is funded by the US National Science Foundation (NSF QII-TAQS) under Grant No.1936246.The authors acknowledge the Rosen Center for Advanced Computing at Purdue University for the use of their computing resources and technical support. We are also grateful for a helpful discussion with Prof. Jay D. Sau, Dr. Tamoghna Barik, Prof. Jennifer E. Hoffman, and Benjamin H. November.

# References

1.      Fanfarillo, L. Go for a spin. *Nat Phys* **18**, 738–739 (2022).

2.      Liu, Y. *et al.* Pair density wave state in a monolayer high-Tc iron-based superconductor. *Nature* **618**, 934–939 (2023).

3.      Fernandes, R. M. *et al.* Iron pnictides and chalcogenides: a new paradigm for superconductivity. *Nature* **601**, 35–44 (2022).

4.  Wang, D. *et al.* Evidence for Majorana bound states in an iron-based superconductor. *Science (1979)* **362**, 333–335 (2018).

5.  Zhang, P. *et al.* Observation of topological superconductivity on the surface of an iron-based superconductor. *Science (1979)* **360**, 182–186 (2018).

6.  Ambolode, L. C. C. *et al.* Se content $x$ dependence of electron correlation strength in Fe$_{1+y}$Te$_{1-x}$Se$_{x}$. (2015) doi:10.1103/PhysRevB.92.035104.

7.  Wang, Z. *et al.* Topological nature of the FeSe0.5Te0.5 superconductor. *Phys Rev B Condens Matter Mater Phys* **92**, 1–7 (2015).

8.  Li, Y. *et al.* Electronic properties of the bulk and surface states of Fe1+yTe1−xSex. *Nat Mater* **20**, 1221–1227 (2021).

9.  Machida, T. *et al.* Zero-energy vortex bound state in the superconducting topological surface state of Fe(Se,Te). *Nat Mater* **18**, 811–815 (2019).

10. Chiu, C. K., Machida, T., Huang, Y., Hanaguri, T. & Zhang, F. C. Scalable Majorana vortex modes in iron-based superconductors. *Sci Adv* **6**, 1–11 (2020).

11. Ghazaryan, A., Lopes, P. L. S., Hosur, P., Gilbert, M. J. & Ghaemi, P. Effect of Zeeman coupling on the Majorana vortex modes in iron-based topological superconductors. *Phys Rev B* **101**, 20504 (2020).

12. Barik, T. & Sau, J. D. Signatures of nontopological patches on the surface of topological insulators. *Phys Rev B* **105**, 1–7 (2022).

13. Zhao, H. *et al.* Nematic transition and nanoscale suppression of superconductivity in Fe(Te,Se). *Nat Phys* **17**, 903–908 (2021).

14. Terao, K., Kashiwagi, T., Shizu, T., Klemm, R. A. & Kadowaki, K. Superconducting and tetragonal-to-orthorhombic transitions in single crystals of FeSe1-xTex (0≤ x ≤0.61) Superconducting and ... KOtaro Terao et al. *Phys Rev B* **100**, 2–8 (2019).

15. Nakayama, K. *et al.* Orbital mixing at the onset of high-temperature superconductivity in FeSe1-xTex/CaF2. *Phys Rev Res* **3**, 1–6 (2021).

16. Lohani, H. *et al.* Band inversion and topology of the bulk electronic structure in FeSe0.45Te0.55. *Phys Rev B* **101**, 1–10 (2020).

17. Aichhorn, M., Biermann, S., Miyake, T., Georges, A. & Imada, M. Theoretical evidence for strong correlations and incoherent metallic state in FeSe. *Phys Rev B Condens Matter Mater Phys* **82**, 1–5 (2010).

18. Xu, G., Lian, B., Tang, P., Qi, X. L. & Zhang, S. C. Topological Superconductivity on the Surface of Fe-Based Superconductors. *Phys Rev Lett* **117**, 1–5 (2016).

19. Sales, B. C. *et al.* Bulk superconductivity at 14 K in single crystals of Fe1+yTexSe1-x. *Phys Rev B* **79**, 094521 (2009).

20. Mizuguchi, Y., Tomioka, F., Tsuda, S., Yamaguchi, T. & Takano, Y. Substitution Effects on FeSe Superconductor. *J Physical Soc Japan* **78**, 074712 (2009).

21. Martinelli, a. From antiferromagnetism to superconductivity in Fe_{1 y}Te_{1. *Phys Rev B* **1**, 211 (2010).

22. Kawasaki, Y. *et al.* Phase diagram and oxygen annealing effect of FeTe 1-xSe x iron-based superconductor. *Solid State Commun* **152**, 1135–1138 (2012).

23.     Perdew, J. P., Burke, K. & Ernzerhof, M. Generalized Gradient Approximation Made Simple. *Phys Rev Lett* **77**, 3865–3868 (1996).

24.     Kresse, G. & Furthmüller, J. Efficient iterative schemes for ab initio total-energy calculations using a plane-wave basis set. *Phys Rev B* **54**, 11169–11186 (1996).

25.     Kresse, G. & Joubert, D. From ultrasoft pseudopotentials to the projector augmented-wave method. *Phys Rev B* **59**, 1758–1775 (1999).

26.     Zhang, Y. & Yang, W. Comment on ``Generalized Gradient Approximation Made Simple''. *Phys Rev Lett* **80**, 890 (1998).

27.     Grimme, S. Semiempirical GGA-type density functional constructed with a long-range dispersion correction. *J Comput Chem* **27**, 1787–1799 (2006).

28.     Pizzi, G. *et al.* Wannier90 as a community code: new features and applications. *Journal of Physics: Condensed Matter* **32**, 165902 (2020).

# Supplemental Materials

The band structures of FeTe$_{1-x}$Se$_x$ ($x$ = 0.0, 0.125, 0.25, 0.375, 0.5, 0.625, 0.75, 0.875, 1.0) calculated by different optimization methods and lattice structures (see Fig. S1-S2) show similar nonlinear relation of SOC gap with respect to $x$.

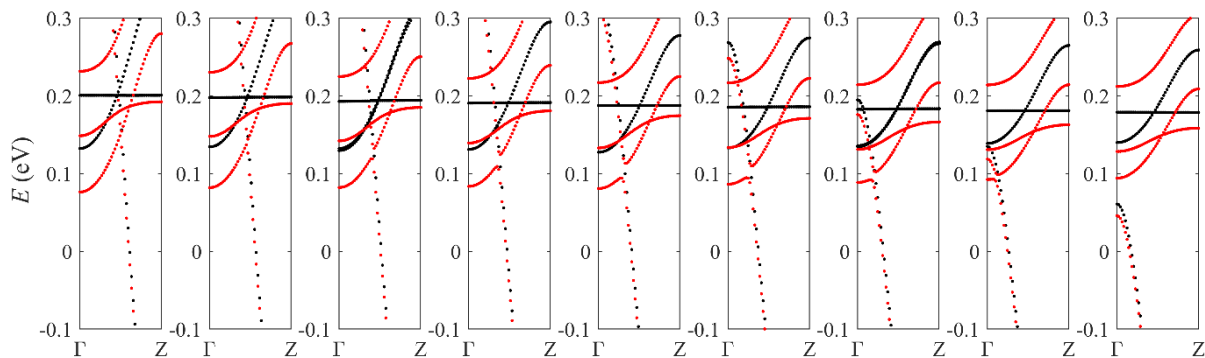

Figure S1. Band structures with (red dot) and without (black dot) considering SOC effect of FeTe$_{1-x}$Se$_x$ ($x$ = 0.0, 0.125, 0.25, 0.375, 0.5, 0.625, 0.75, 0.875, 1.0) calculated by DFT based on FeTe$_{0.5}$Se$_{0.5}$[] crystal structure without optimization. The lattice constants $a$ = $b$ =7.9466 (Å) and $c$ = 5.9552 (Å), and Se and Te atoms are at the same plane.

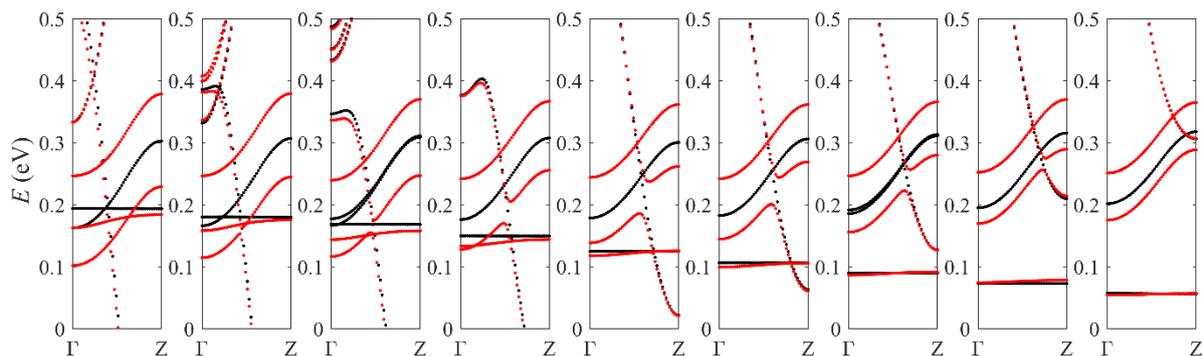

Figure S2. Band structures with (red dot) and without (black dot) considering SOC effect of FeTe$_{1-x}$Se$_x$ ($x$ = 0.0, 0.125, 0.25, 0.375, 0.5, 0.625, 0.75, 0.875, 1.0) calculated by DFT. The lattice constants are set to be $a$ = 7.6406 - 0.108$x$ (Å), $c$ = 6.3362 - 0.762$x$ (Å), and atomic geometries are optimized using PW86R exchange[] and PBE correlation[] with nonlocal vdW-DF2 functional[].

Table S1. DFT calculated SOC gap $\Delta$ and p-d SOC coupling parameter $\lambda_3$ in pd effective Hamiltonian for

FeTe$_{1-x}$Se$_x$ systems.

| $x$ | 0 | 0.125 | 0.25 | 0.375 | 0.5 | 0.625 | 0.75 | 0.875 | 1 |
|---|---|---|---|---|---|---|---|---|---|
| $\lambda_3$ (meV) | 0.0 | 3.2 | 4.7 | 8.9 | 15.7 | 14.3 | 13.6 | 0.1 | 0.0 |
| $\Delta$ (meV) | 0.0 | 9.4 | 19.0 | 33.4 | 48.2 | 41.0 | 32.9 | 0.0 | 0.0 |

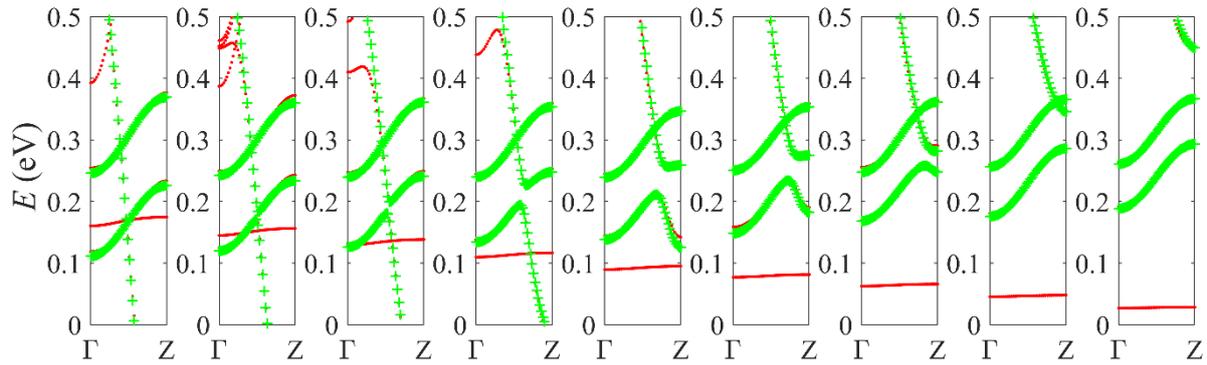

Figure S3. Band structures of FeTe$_{1-x}$Se$_x$ ($x$ = 0.0, 0.125, 0.25, 0.375, 0.5, 0.625, 0.75, 0.875, 1.0) calculated

by DFT (red dot) and fitted by effective Hamiltonian (green cross) with considering SOC effect.

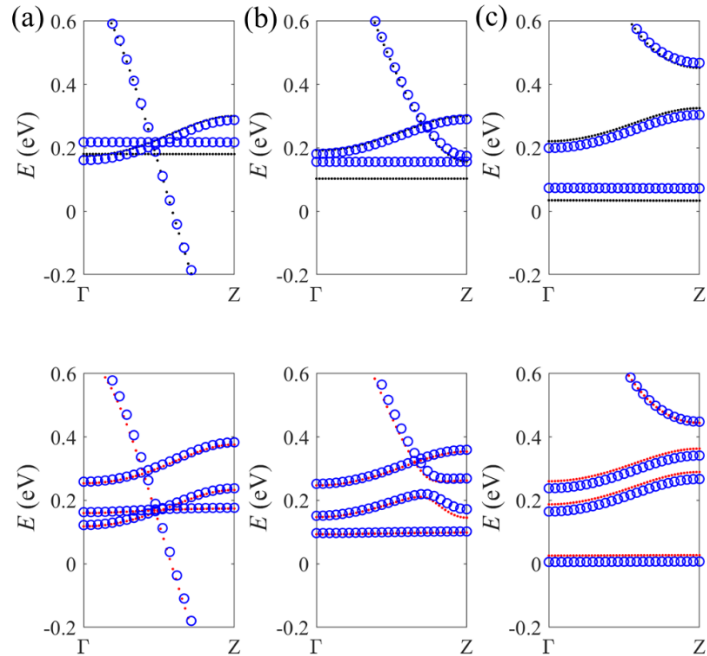

Figure S4. Band structures of FeTe$_{1-x}$Se$_x$ with $x$ = 0.0 (a), 0.5 (b), 1.0 (c), calculated by DFT without (black

dot) or with SOC effect (red dot) and corresponding Wannier-based fitting results (blue circle).

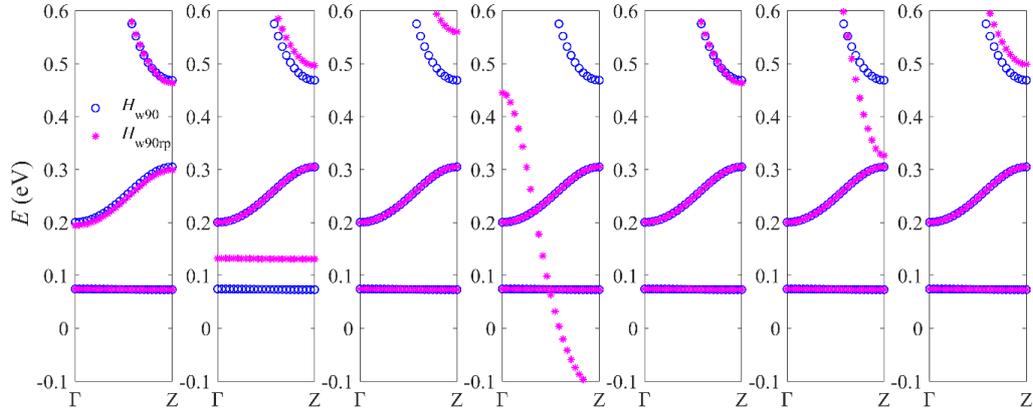

Figure S5. Band structures of FeSe$_x$Te$_{1-x}$ with $x = 1.0$ calculated by Wannier Hamiltonian without SOC effect (blue circle, $H_{w90}$) and the bands (magenta star, $H_{w90rp}$) calculated by this Hamiltonian with $\varepsilon_{d_{xz/yz}}$, $\varepsilon_{d_{x^2-y^2}}$, $\varepsilon_{p_z}$, $|t_{dx^2-y^2 \sim p_z}|$, $|t_{dxz/yz \sim p_z}|$, $t^0_{pz \sim pz}$, $t^1_{pz \sim pz}$ (from left to right) replaced by the corresponding terms in case of $x = 0.0$, respectively.